\documentclass[journal, 10pt]{IEEEtran}

\newcommand{\eref}[1]{(\ref{#1})}
\usepackage{amsthm} 

\newtheorem{thm}{Theorem}

\newtheorem{lemma}[thm]{Lemma}

\newtheorem{example}{Example}
\newtheorem{proposition}{Proposition}
\newtheorem{definition}[thm]{Definition}

\newtheorem{remark}{Remark}
\usepackage{amsmath,amssymb}
\usepackage{mathtools}
\usepackage{cite}
\usepackage{mathrsfs}
\usepackage{subfig}
\usepackage{array}
\usepackage{stfloats}
\usepackage{color}
\usepackage{todonotes}
\usepackage{multirow}
\usepackage{hhline}
\usepackage{epsf, pstricks,pgf,xcolor}
\usepackage{epstopdf,epsfig}
\graphicspath{{figs/}}

\title{
Non-Stationary Polar Codes for Resistive Memories
}
\author{Marwen Zorgui, Mohammed E. Fouda, Zhiying Wang, Ahmed M. Eltawil and Fadi Kurdahi\\
Electrical Engineering and Computer Science, University of California--Irvine, CA, USA}
\begin{document}

\maketitle
\thispagestyle{empty}
\pagestyle{empty}
\begin{abstract}
Resistive memories are considered a promising memory technology enabling high storage densities with in-memory computing capabilities. 
However, the readout reliability of resistive memories is impaired due to the inevitable existence of wire resistance, resulting in the sneak path problem. 
Motivated by this problem, we study polar coding over channels with different reliability levels, termed non-stationary polar codes, and we propose a technique improving its bit error rate (BER) performance. We then apply the framework of non-stationary polar codes to the crossbar array and evaluate its BER performance under two modeling approaches, namely binary symmetric channels (BSCs) and binary asymmetric channels (BSCs). Finally, we propose a technique for biasing the proportion of high-resistance states in the crossbar array and show its advantage in reducing further the BER. Several simulations are carried out using a SPICE-like simulator, exhibiting significant reduction in BER. 
\end{abstract}

\section{INTRODUCTION}
Emerging nonvolatile memories (NVMs), such as phase change memory (PCRAM), ferroelectric memory (FeRAM), spin transfer torque magnetic memory (STT-MRAM), and resistive memory (RRAM), have shown high potential as alternatives for floating-gate-based nonvolatile memories \cite{daly2017through}. RRAMs are considered the best candidate for the next generation nonvolatile memory due to their high reliability, fast access speed, multilevel capabilities and stack-ability creating 3D memory architectures. To achieve higher density, 
access devices such as transistors, diodes and selectors are removed. However, the main disadvantage of the selector-less (gate-less) crossbar-based memories is the sneak path effects which limit the readability of the array. 

\begin{figure} 
\centering
\includegraphics[width=0.68\linewidth]{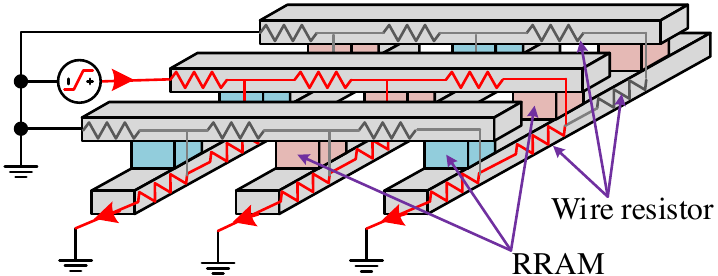}
\label{Fig2b}
\caption{Parallel reading of the entire row in the crossbar. The columns and rows are grounded, except the row being read. The red arrow shows the sensed current flowing through wire resistances and RRAMs.}
\label{Fig2}
\end{figure}

\begin{figure}
\centering
\subfloat[]{\includegraphics[width=0.55\linewidth]{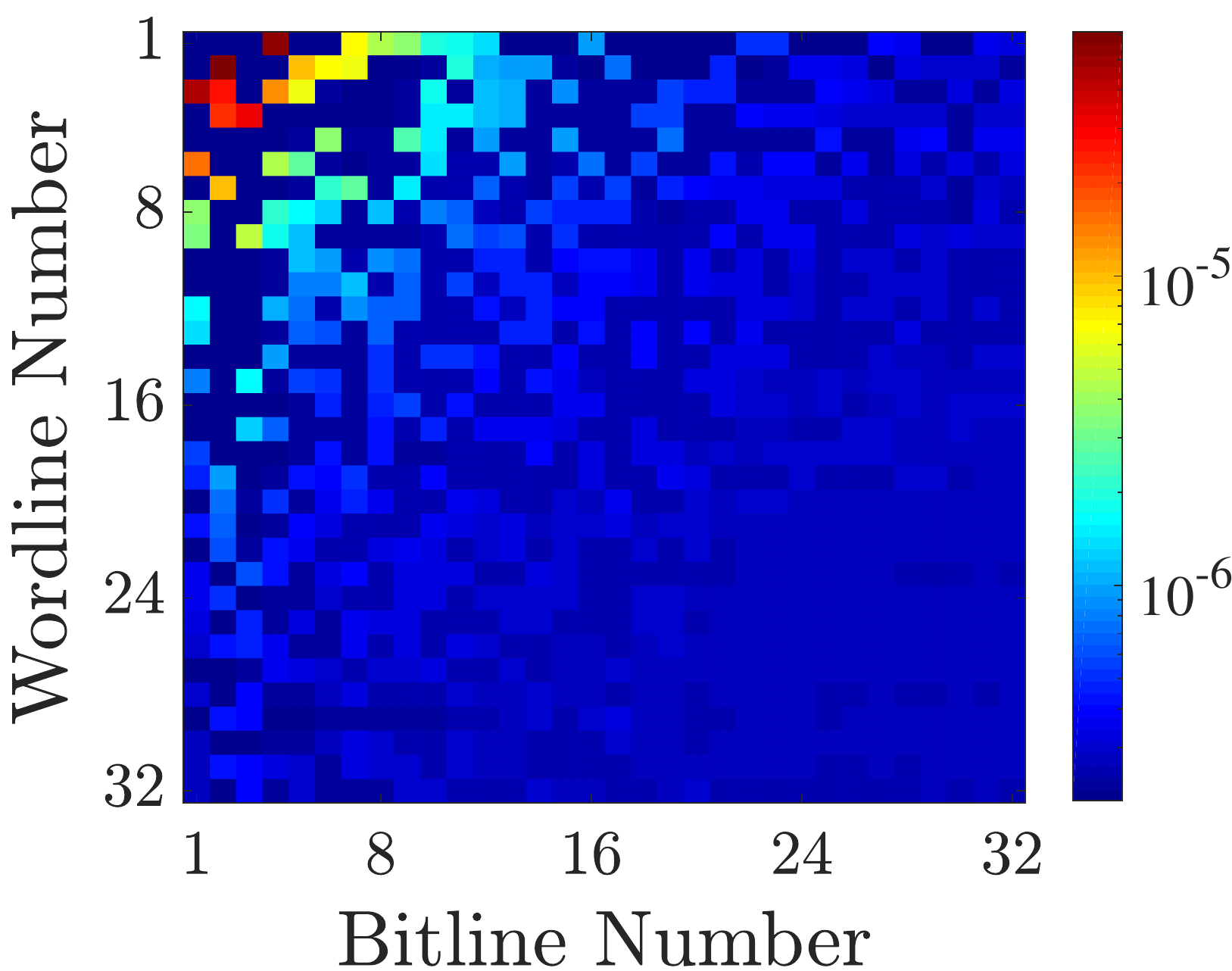}%
\label{Fig3a}}
\hfill
 \subfloat[]{\includegraphics[width=0.32\linewidth,height=0.26\linewidth]{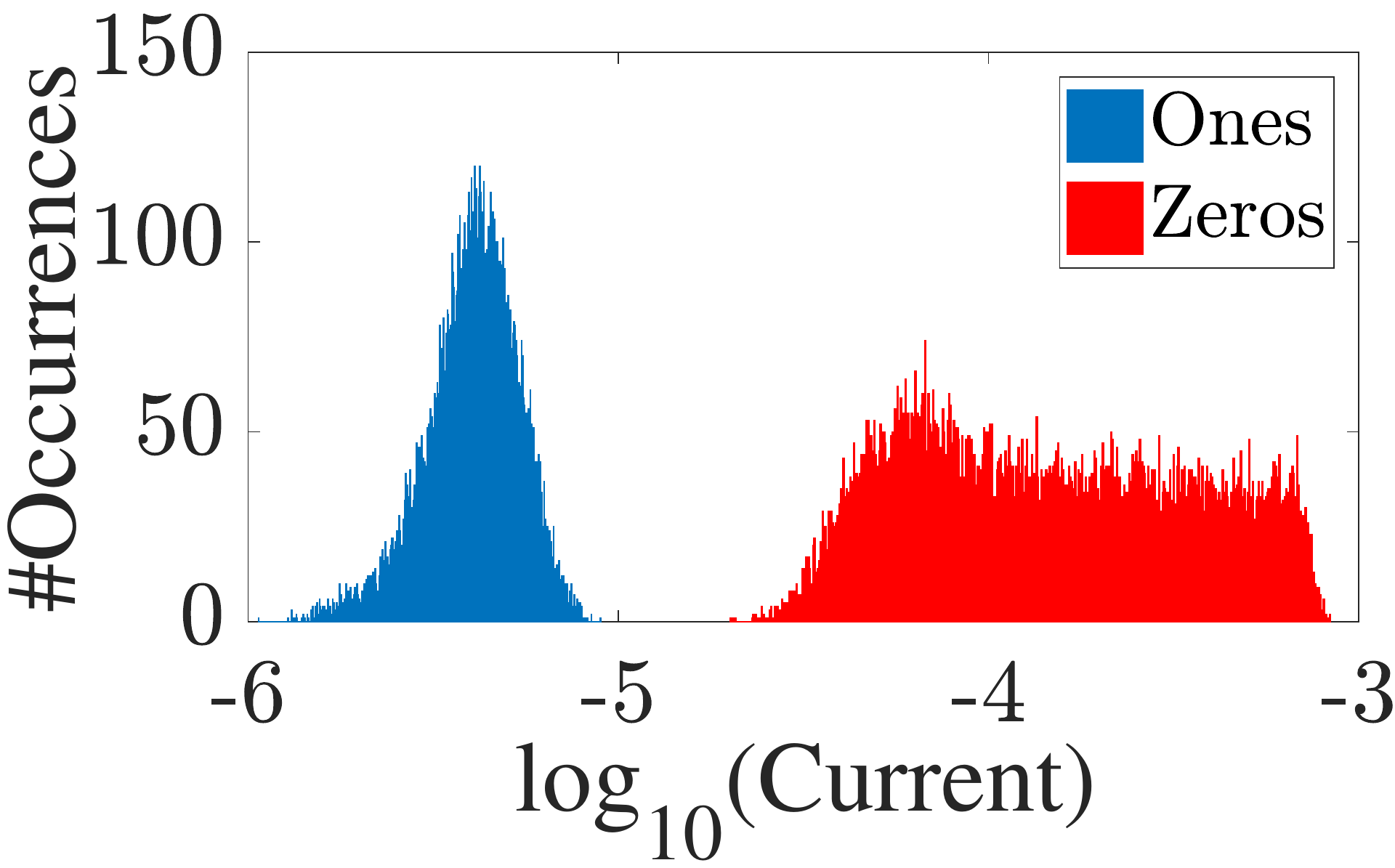}%
 \label{Fig3b}}
\subfloat[]{\includegraphics[width=0.32\linewidth,height=0.25\linewidth]{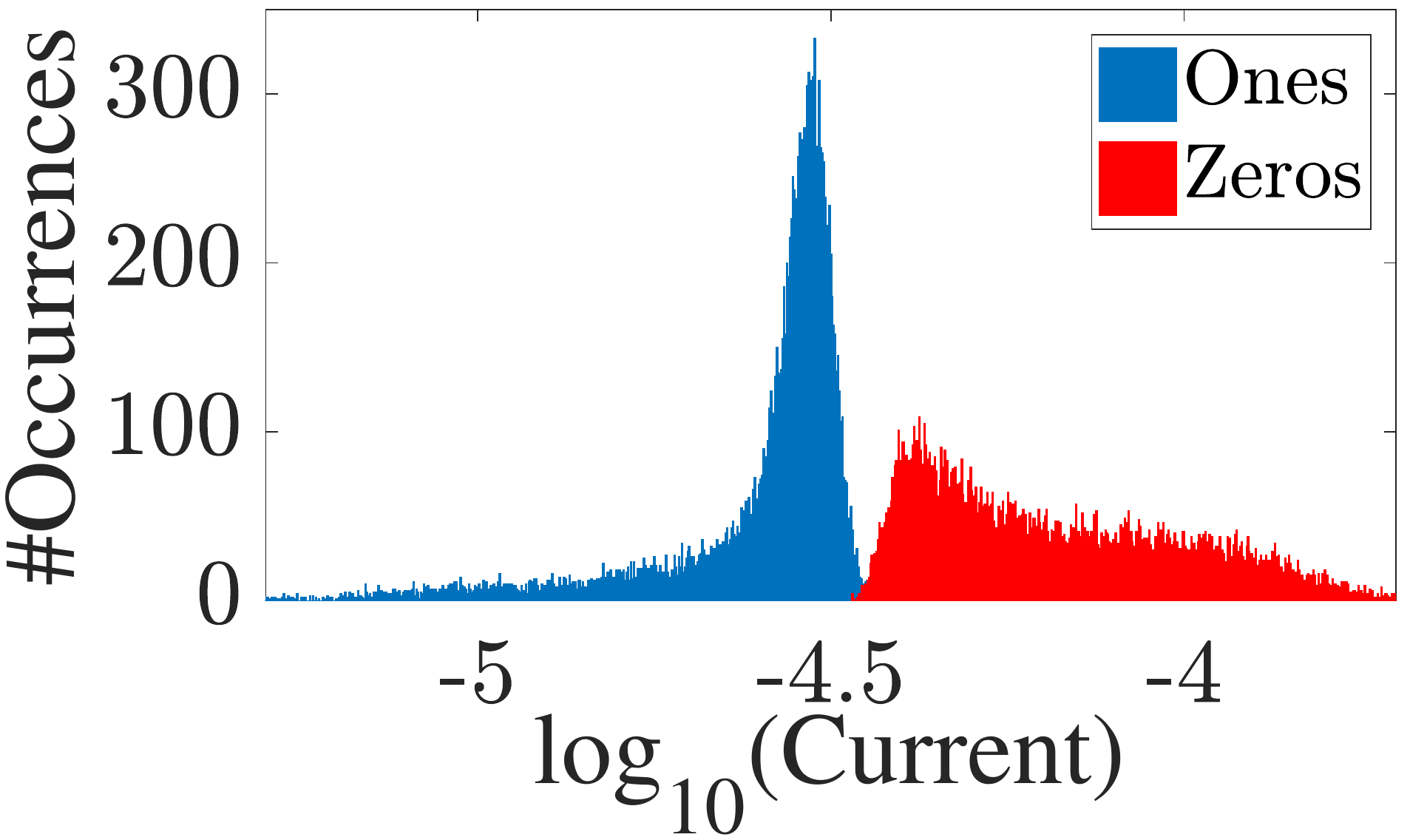}%
\label{Fig3c}}
\subfloat[]{\includegraphics[width=0.32\linewidth,height=0.25\linewidth]{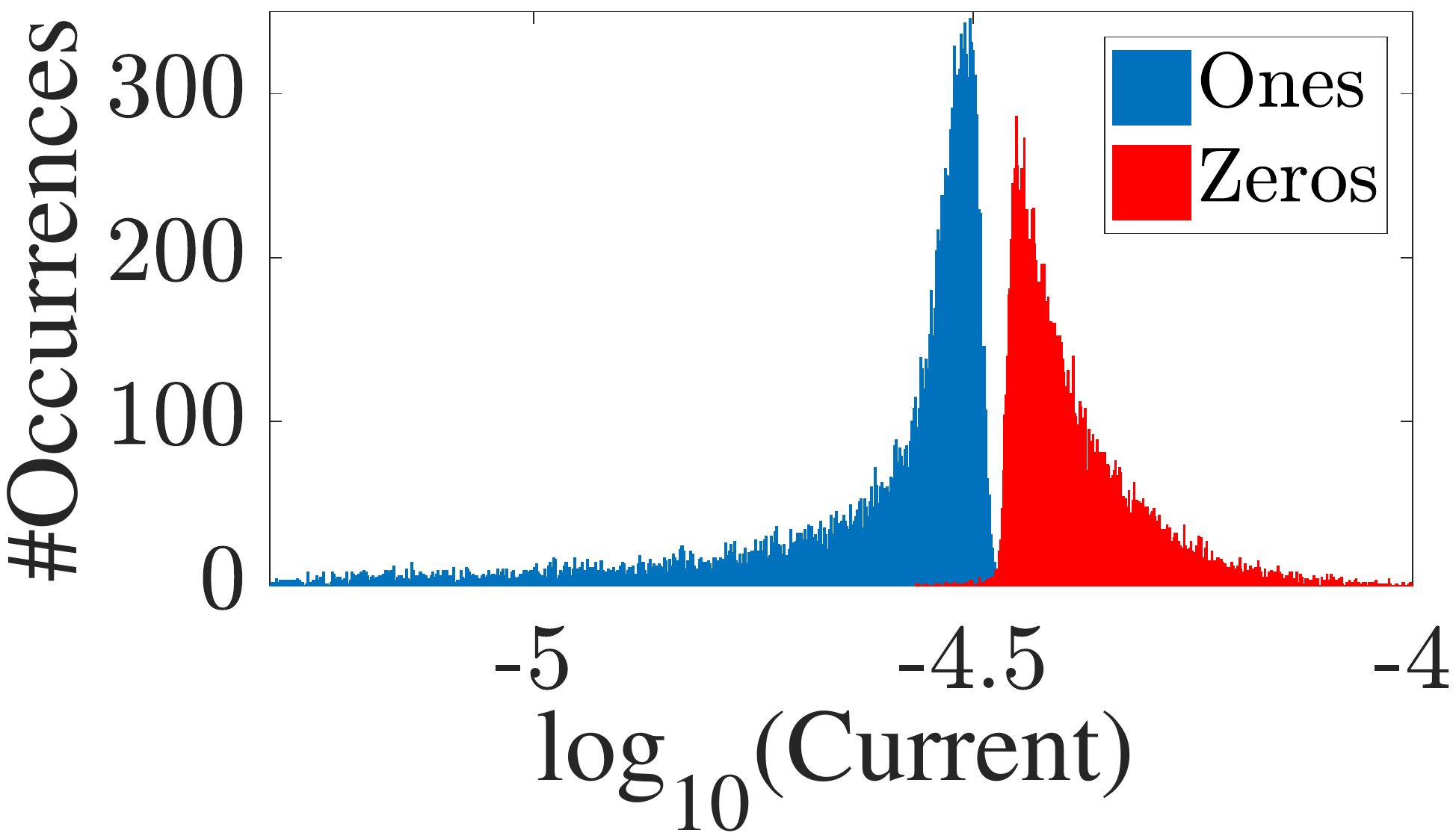}%
\label{Fig3d}}
\caption{(a) Measured current per cell, and its histogram for bitline number (b) $1$, (c) $16$, and (d) $32$.}
\label{Fig3}
\end{figure}

{

In resistive memories, the high/low cell resistance represents the stored bit. In order to retrieve the stored data, resistive sensing (reading) techniques are adopted.
In this paper, a parallel reading of an entire crossbar row \cite{fouda2019resistive} is adopted (see Fig. \ref{Fig2}). It eliminates the multi-path problem in single-cell reading \cite{zidan2016single}, one of the causes of sneak path. But, the inevitable wire resistances lead to undesirable voltage drops, another type of sneak path problem. These voltage drops are functions of the stored data and the wire resistance. At the expected feature size of $F=5nm$ of RRAMs, the wire resistance per cell reaches as high as $90 \Omega$~\cite{fouda2018modeling}.}
Fig.~\ref{Fig3a} shows the measured current of each cell in a $(32~\times~32)$ array  with $25\Omega$ wire resistance, storing random data, which is generated by the SPICE-like simulator of \cite{fouda2018modeling}. Due to sneak path, the sensed current of low resistance state decreases in both vertical and horizontal directions in the array. The top left cells have less disturbed behavior. The stored bits are distinguishable. On the other hand, bits in the right-bottom cells are indistinguishable due to the read margin overlap. Fig.~\ref{Fig3b}, Fig.~\ref{Fig3c} and Fig.~\ref{Fig3d} show the histogram of the measured currents of the $1$st, $16${th} and the $32$nd bitline, respectively. Clearly, the larger the bitline index is, the more errors occur. We can see that the channels of the cells have varying reliability.
Addressing the sneak path problem has attracted a lot of interest from both research and industry communities. Proposed solutions include hardware-based approaches, e.g., transistor gating~\cite{zidan2013memristor}, and/or techniques based on communication and coding theory~\cite{chen2018coding, ben2019detection}. The focus of this paper is on the latter approach and our scheme is based on polar codes. Polar codes~\cite{arikan2009channel} are the first family of explicit error-correcting codes to provably achieve the capacity of binary symmetric channels, with a low-complexity encoding and successive cancellation decoding (SCD). For a code of length $N$, encoding/decoding has a complexity of $O(N \log N )$. For the above reasons, polar codes constitute an attractive error correction scheme. 

Our contributions are summarized as follows:
\begin{itemize}
    \item We study polar coding for channels with different reliability levels, termed non-stationary polar codes. 
    \item We argue that the ordering of the channels is important and propose an ordering
with a competitive performance numerically.
 \item	We then apply the framework of non-stationary polar codes to the sneak path problem and demonstrate significant improvement in terms of bit error rate (BER). In particular, we discuss modeling the array cells as either binary symmetric channels (BSCs) or as binary asymmetric channels (BACs) and compare their BER performances. 

\item	Finally, we propose a
technique for biasing the number of high-resistance values in
the crossbar through puncturing, and show by simulation that the proposed technique
can help reduce the BER in certain scenarios.
\end{itemize}

The remainder of the paper is organized as follows. In Section~\ref{polar_code_framework}, we analyze the framework for polar codes over channels with different reliability levels and describe our approach for employing such polar codes. In Section~\ref{General_framework}, we apply the developed framework to the crossbar array and describe the various techniques proposed to further reduce the BER. Section~\ref{concluding_remarks} draws conclusions and discusses future research directions.

\noindent\textbf{Notation}: Vectors are denoted with lower-case bold letters. A permutation $\pi$ over the integers $\{0, \ldots, N-1 \}$ is denoted as $\pi = [ \pi(0), \ldots, \pi(N-1) ]$, where $\pi(j)$ is the image of $j$ under $\pi$. For a vector $\mathbf{x}=[x_0, \ldots, x_{N-1}]$, $\mathbf{x}_{\pi}$ denotes the vector $\mathbf{x}_{\pi}= [x_{\pi(0)}, \ldots, x_{\pi(N-1)}]$.
\section{Non-stationary polar code construction}
\label{polar_code_framework}
For a binary input discrete memoryless channel (B-DMC), $W$, with output alphabet $\mathcal{Y}$, we denote its transition probabilities by $W(y|x), x \in \{0,1 \}, y \in \mathcal{Y}$, and define the symmetric channel output probability as $W(y)=\frac{1}{2}W(y|0)+ \frac{1}{2}W(y|1)$.
Define the symmetric capacity $I(W)$ as

\vspace{-0.3cm}
\begin{small}
\begin{align}
\label{symmetric_capacity}
I(W)   \triangleq  \sum\limits_{y \in \mathcal{Y}} \sum\limits_{x \in \{0,1 \}} 
\frac{1}{2} W(y|x) \log \frac{W(y|x)}{W(y)}.
\end{align}
\end{small}

\vspace{-0.3cm}
\noindent	Polar codes \cite{arikan2009channel} manufacture out of $N$ independent copies of a given B-DMC channel, a second set of $N$ \emph{synthesized} binary-input channels.
The channels show a polarization effect, namely, as $N$ becomes large, the symmetric capacities of the synthesized channels tend towards 0 or 1 for all but a vanishing fraction. 

\begin{figure} 
\centering
\vspace{-0.1in}
\includegraphics[width=.7\linewidth ]{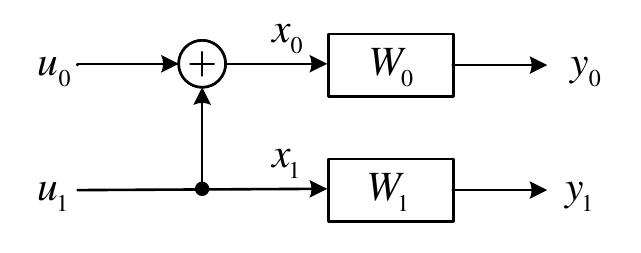}
\caption{Basic channel polarization.}
\vspace{-0.1in}
\label{Basic_polarization}
\end{figure}

In our framework, in contrast to the original polar codes, 
we consider the extension of polar codes to the setting where the underlying channels are of \emph{varying reliability levels} 
(see Fig.~\ref{Basic_polarization} and Fig.~\ref{PCdiagram}). 
Using the terminology in \cite{alsan2016simple}, we refer to such polar codes as non-stationary polar codes. 

The basic polarization transformation, illustrated in Fig.~\ref{Basic_polarization}, is applied to two independent channels $W_0: \{ 0,1\} \to \mathcal{Y}_0$ and $W_1: \{ 0,1\} \to \mathcal{Y}_1$, resulting in two channels, $W': \{0,1\} \to \mathcal{Y}_0 \times \mathcal{Y}_1$ and $W'': \{0,1\} \to \mathcal{Y}_0 \times \mathcal{Y}_1\times \{0,1\}$, given by
\begin{equation}
\begin{aligned}
W'(y_0, y_1 | u_0) = \frac{1}{2} \sum\limits_{u_1 \in \{0,1 \}} W_0( y_0 | u_0 \oplus u_1)  W_1( y_1 | u_1), \\
W''(y_0, y_1, u_0 | u_1) = \frac{1}{2}  W_0( y_0 | u_0 \oplus u_1)  W_1( y_1 | u_1),
\label{single_step_polarization}
\end{aligned}
\end{equation}
\noindent where $y_0 \in \mathcal{Y}_0, y_1 \in \mathcal{Y}_1$, and $u_0,u_1 \in \{ 0,1\}$. We denote this single-step transformation by $(W_0, W_1)  \mapsto (W', W'' )$. 
The transformation preserves the average symmetric capacity, while exhibiting a polarization effect.
\begin{lemma}[\cite{alsan2016simple}]
Suppose $(W_0, W_1)  \mapsto (W', W'' )$. Then, 
\begin{equation}
\begin{aligned}
\label{preserve}
&I(W') + I(W'')  =  I(W_0) + I(W_1), \\
& I(W') \le  I(W_i)   \le I(W''), i =0,1.
\end{aligned}
\end{equation}
\end{lemma}
The Bhattacharya parameter of a binary discrete memory-less channel $W: \{0,1 \} \to \mathcal{Y}$, denoted by $Z(W)$ is a measure of the reliability of $W$, and has been used to bound the error probability of polar codes in \cite{arikan2009channel}. The parameter $Z(W)$ is defined as
\begin{align*}
Z(W) \triangleq \sum_{y \in \mathcal{Y}} \sqrt{W(y|0) W(y|1)}.
\end{align*}

\begin{lemma} 
\label{single_step_results}
Let $W_0: \{ 0,1\} \to \mathcal{Y}_0$ and $W_1: \{ 0,1\} \to \mathcal{Y}_1$ and $(W_0, W_1)  \mapsto (W', W'' )$.\\
(i) The following relations hold:
\begin{align}
\label{polarization_output_1}
Z( W') &\le Z(W_0) +  Z(W_1) - Z(W_0) Z(W_1), \\
\label{polarization_output_2}
Z(W'') &= Z(W_0) Z(W_1), \\
Z(W') &\geq \sqrt{Z(W_0)^2 + Z(W_1)^2 - Z(W_0)^2 Z(W_1)^2 }.
\label{polarization_lowed_bound}
\end{align}
(ii) Equality ~\eref{polarization_lowed_bound} holds with equality iff both $W_0$ and $W_1$ are binary symmetric channels.\\
(iii) Equality ~\eref{polarization_output_1} holds with equality iff $W_0$ or $W_1$ is a binary erasure channel. 
\end{lemma}
In Lemma \ref{single_step_results}, parts (i) and (ii) are from \cite[Lemma 2.15 and Lemma 2.16]{korada2009polar} and \cite[Lemma 9]{mahdavifar2017fast}, and part (iii) can be proved following similar lines as~\cite[Proposition 5]{arikan2009channel}.
Lemma~\ref{single_step_results} implies that reliability improves under a single-step channel transformation in the sense that 
\begin{align*}
Z(W') + Z( W'') \le Z(W_0) + Z(W_1),
\end{align*} 
\noindent with equality iff $W_0$ or $W_1$ is a binary erasure channel (BEC).

The single-step transformation, as described in \eref{single_step_polarization}, can be generalized to the case of $N=2^n$ binary, memory-less, but not necessarily stationary channels $\{W_i \}_{i=0}^{N-1}$. In particular, let $W_{m,i}$ denote the $i$-th bit channel after $m$ levels of polarization of the sequence $\{W_i \}_{i=0}^{N-1}$, where $ W_{0,i} \triangleq
W_{ i}, i = 0, \ldots,N-1$. Applying Arikan's polar construction results in a collection of synthesized channels, such that, for any level $1\le l \le n$, for $0 \le j < 2^{l-1}, 0 \le  m <  2^{n-l}$, we have
\begin{align}
&(W_{l-1, 2^l m + j}, W_{l-1, 2^l m + 2^{l-1}+ j})\mapsto   \nonumber\\ 
& \qquad(W_{l, 2^l m + j}, W_{l, 2^l m + 2^{l-1}+ j}).
\label{general_formula}
\end{align}
In \cite{alsan2016simple}, it was shown that the fraction of non-polarized channels approach 0, i.e., for every $0 < a < b < 1$, 
\begin{align*}
\liminf_{N \to \infty} \frac{1}{N} | 0 \le i < N: I(W_{n,i}) \in [a,b]  \}|= 0.
\end{align*} 
The work in~\cite{mahdavifar2017fast} proved that in the asymptotic regime where the blocklength $N$ is large, the effective average symmetric capacity $ \bar{I} (\{ W_i \}_{i=0}^{\infty}) \triangleq \lim\limits_{N \to \infty} \frac{1}{N} \sum_{i = 1} ^{N} I(W_i)$ is achievable. Moreover, the polar coding scheme is constructed based on Arikan's channel polarization transformation in combination with certain permutations at each polarization level. However, it is not clear whether such permutation choices yield attractive performances for practical lengths. In this work, we are concerned with the performance of non-stationary polar codes in the finite blocklength regime. 
\subsection{Polar Code Encoding and Decoding}
Let $\mathbf{u} = [u_0, u_1, \ldots, u_{N-1}]$ and $\mathbf{x} = [x_0, x_1, \ldots, x_{N-1}]$ be the input and the output of a length-$N$ polar code, respectively, with $N=2^n$ for some integer $n$. The encoding of polar codes is given by 
\begin{align*}
 \mathbf{x} = \mathbf{u} G ,  \quad 
 G  = \begin{bmatrix}
 1 & 0\\
 1 & 1
 \end{bmatrix}^{\otimes n},
\end{align*}
where the symbol $^{\otimes n}$ denotes the $n$-th Kronecker power operator.
%
After polar encoding, one obtains $N$ synthesized channels $\{ W^{(i)} \triangleq W_{n,i} , 0 \le i \le N-1 \}$.
A polar code of dimension $k$ transmits $k$ information bits in the $k$ synthesized channels with the highest $I(W^{(i)})$ (we denote the corresponding information set by $\mathcal{I}$), and $N-k$ arbitrary but fixed bits in the remaining $N-k$ synthesized channels (denoted by $\mathcal{F}$). Decoding of polar codes is carried out using successive cancellation decoding (SCD) as in \cite{arikan2009channel}, taking into account the appropriate likelihood ratios of the original channels.
\begin{remark}
\label{computing_information_set}
The information set of a non-stationary polar code can be determined as long as $I(W^{(i)})$ are computed. However, their computational complexities are generally unmanageable. Instead, we propose that the reliability parameters of the synthesized channels can be approximated efficiently by $Z(W^{(i)})$ using \eref{polarization_output_1} and \eref{polarization_output_2} in Lemma~\ref{single_step_results}. We call this method the Bhattacharyya bound approach.
It is a generalization of the algorithm in~\cite{arikan2008performance}, developed for regular polar codes. Let $Z_{n,i}$ denote the Bhattacharyya parameter of channel $W_{n,i}$. Then, we apply the following recursion: for level $1 \le l\le n$, for $0 \le j < 2^{l-1}$ and $0 \le m < 2^{n-l}$, we have 
\begin{align*}
 Z_{l, 2^l m + j} &= Z_{l-1, 2^l m + j}+ Z_{l-1, 2^l m + 2^{l-1}+ j} \nonumber\\
 &\quad	- Z_{l-1, 2^l m + j}  Z_{l-1, 2^l m + 2^{l-1}+ j}, \nonumber\\
 Z_{l, 2^l m + 2^{l-1}+ j} &= Z_{l-1, 2^l m + j}  Z_{l-1, 2^l m + 2^{l-1}+ j},
\end{align*}
where $\{Z_{0,i}, 0 \le i < N\}$ are the initial channel Bhattacharyya parameters. The indices of the lowest $N-k$ values in the set of $N$ final stage values form the set $\mathcal{F}$. This algorithm is an evolution of the Bhattacharyya parameters of channels from right to left (see Fig. \ref{PCdiagram}), preferably applied in log-domain to avoid underflow. 
\end{remark}
\subsection{Role of the Channel Ordering}
\begin{figure} 
\centering
\includegraphics[width=\linewidth ]{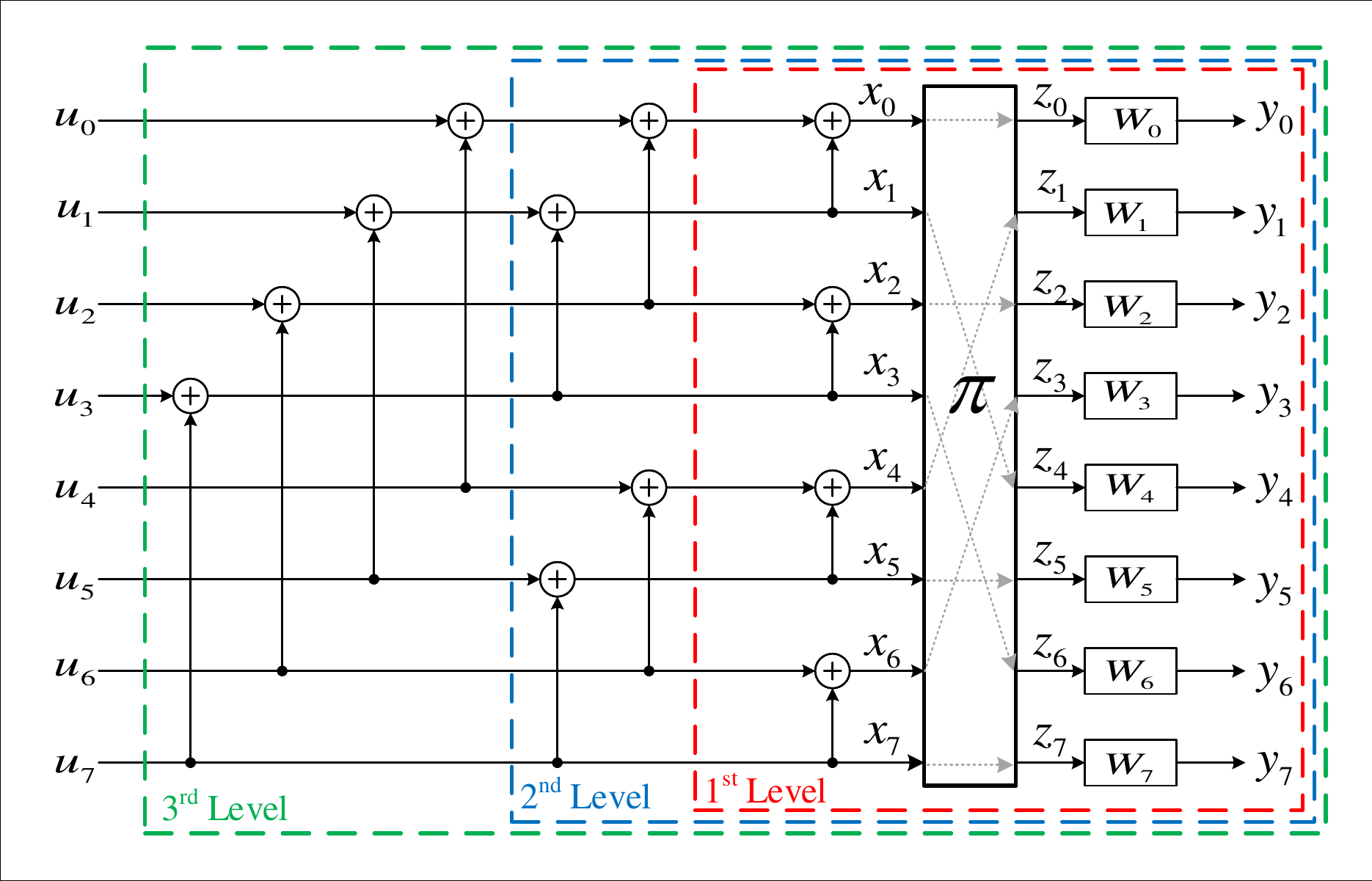}
\caption{Polar encoding with $N=8$ channels, with permutation $\pi = [0,4,2,6,1,5,3,7]$.} 
\label{PCdiagram}
\end{figure}

The performance of polar codes depends on the synthesized channels of the information set $\sum_{i \in \mathcal{I}} I(W^{(i)})$~\cite{arikan2009channel}. As illustrated in Fig.~\ref{PCdiagram}, to construct a non-stationary polar code, we propose to apply a permutation $\pi$ to the vector $\mathbf{x}$ in order to enhance the overall performance. Ideally, we want to find a permutation $\pi$ such that 
\begin{align}
\sum_{i \in \mathcal{I}_{\pi^*}} I(W^{(i)}_{\pi^*}) \geq \sum_{i \in \mathcal{I}_{\pi }} I(W^{(i)}_{\pi }),
\label{permutation_choice}
\end{align}
where $I( W^{(i)}_{\pi})$ is the symmetric capacity of the $i$-th synthesized channel under permutation $\pi$ and $\mathcal{I}_{\pi}$, its information set. 
\begin{remark}
The permutation $\pi$ is defined such that $\mathbf{z}_\pi  =\mathbf{x}$, or equivalently $\mathbf{z}  =\mathbf{x_{\pi^{-1}}}$. Then, $z_{\pi(i)} = x_i$, implying that symbol $x_i$ goes through channel $W_{\pi(i)}$, for $0 \le i < N$.
Correspondingly, a reverse permutation is required for the output of the channels. The polar decoder receives as input the vector $\mathbf{y}'= [y_{\pi(0)}, \ldots, y_{\pi(N-1)}] = \mathbf{y}_\pi$.
See Fig.~\ref{fullmodel} for a schematic picture of the full encoder and decoder.
\end{remark}
\begin{remark}
\label{general_order}
In general, the channel symmetric capacities can be given in any arbitrary order. In this case, in order to study the effect of permutations in a canonical way, we apply the composition permutation $\pi_{\text{ord}} \circ \pi$ to the output $\mathbf{x}$, where $\pi_{\text{ord}}$ is a permutation that orders the channels in increasing order of symmetric capacities: that is $I( W_{\pi_{\text{ord}} (i)}) \le I( W_{\pi_{\text{ord}} (j)})$, for $i < j$. Thus, the output of the polar encoder $x_i$ is mapped to channel  $ W_{ \pi_{\text{ord}} \circ \pi (i)}, i.e., x_i = z_{\pi_{\text{ord}} \circ \pi (i)}$.
\end{remark}
We next show that some channel orderings are equivalent after polarization.
\begin{definition}
Given $N$ B-DMC channels  $\{W_{i} \}_{i=0}^{N-1}$, an equivalence class of permutations over $\{0, \ldots, N-1 \}$ consists of all channel permutations $\pi$ such that the $n$-level polarization process of the channels $\{W_{\pi(i)} \}_{i=0}^{N-1}$ results in the same symmetric capacities.
\end{definition}
\begin{lemma}
\label{symmetry}
The number of permutation classes is upper bounded by $\frac{N!}{2^{N-1}}$.
\end{lemma}
\begin{IEEEproof}
Consider the two single-step transformations $(W_0, W_1)  \mapsto (W_1', W_1'' )$ and $(W_1, W_0)  \mapsto (W_2', W_2'' )$. 
From~\eref{symmetric_capacity} and \eqref{single_step_polarization}, we have 
\begin{align}
I(W_1'')&=  \sum\limits_{y_0, y_1, u_0} \sum\limits_{u_1} \frac{1}{4}  W_0( y_0 | u_0 \oplus u_1)  W_1( y_1 | u_1) \nonumber\\& \times \log \frac{\frac{1}{2}  W_0( y_0 | u_0 \oplus u_1)  W_1( y_1 | u_1)}{ W_1''(y_0,y_1,u_0)} \nonumber\\
&=\sum\limits_{y_0, y_1, u_0} \sum\limits_{u_1} \frac{1}{2}  W_0( y_0 |  u_1)  W_1( y_1 | u_0 \oplus u_1) \nonumber\\& \times \log \frac{W_0( y_0 |  u_1)  W_1( y_1 | u_0 \oplus u_1)}{ W_1''(y_0,y_1,u_0)},
\label{part_1}
\end{align}
where the second equality is obtained by a change of variable. On the other hand, we have
\begin{align}
W_1''(y_0,y_1,u_0)&= \sum\limits_{u_1 \in \{ 0,1\}} W_0(y_0|  u_0 \oplus u_1 ) W_1(y_1| u_1 )
\nonumber\\
&= \sum\limits_{u_1 \in \{ 0,1\}} W_0(y_0|   u_1 ) W_1(y_1| u_0 \oplus u_1 ) \nonumber\\
&=W_2''(y_0,y_1,u_0).
\label{part_2}
\end{align}
From \eref{part_1} and \eref{part_2}, we obtain $I(W_1'') = I(W_2'')$. From \eref{preserve}, it follows $I(W_1') = I(W_2')$. Thus, the single-step transformation is symmetric in the sense that exchanging the order of $W_0$ and $W_1$ does not impact the symmetric capacities of the synthesized channels.
\begin{figure}
\centering
\includegraphics[width=.9\linewidth ]{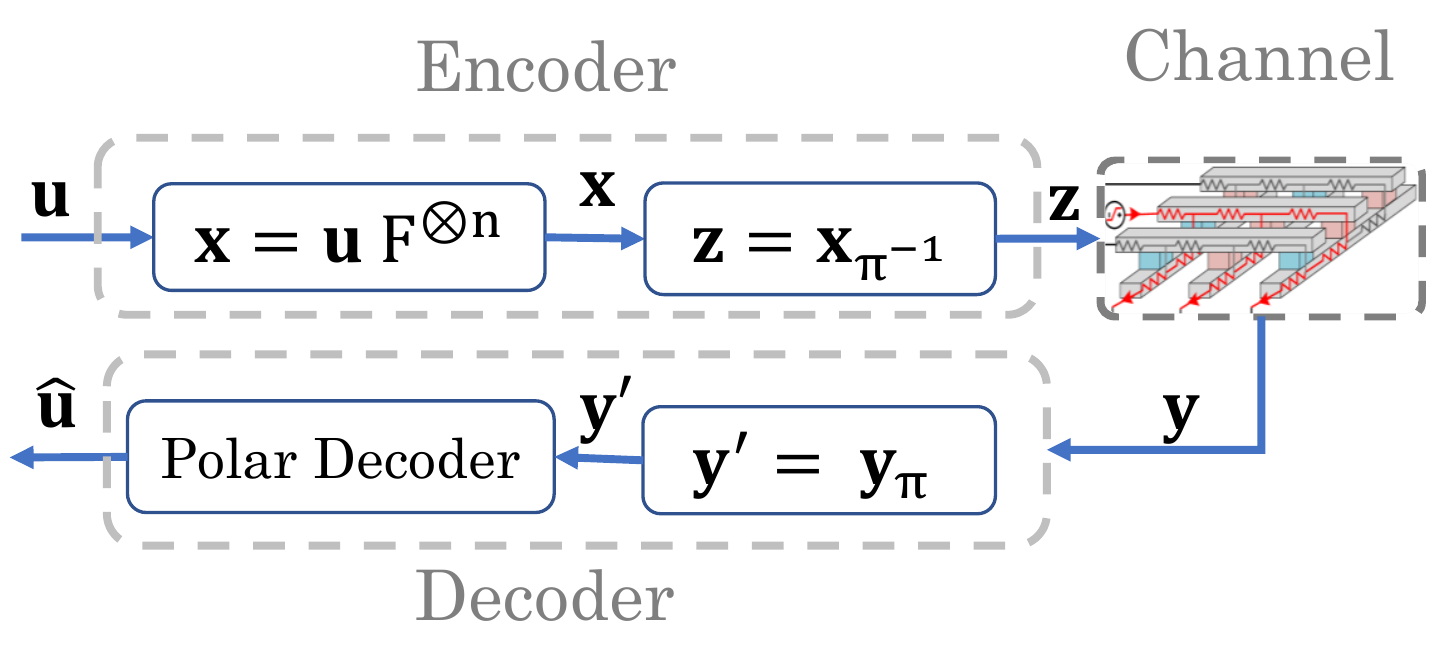}
\caption{Full system model.}
\label{fullmodel}
\end{figure}
It follows that for $N=2^n$ channels, permuting channel $W_{2i}$ and channel $W_{2i+1}$, for $i \in \{0, \ldots, \frac{N}{2}-1 \}$ does not change the symmetric capacities of the synthesized channels obtained at the end of the polarization process. 
Extending the above reasoning to level $2$ in the polarization process, one observes that permuting channels $(W_0, W_1)$ with $(W_2, W_3)$ results in the same symmetric capacities of the synthesized bit channels, and the result holds true for the subsequent pairs of channels in level $2$. Similarly, looking at level $3$, we can also permute $(W_0, W_1,W_2, W_3)$ with $(W_4, W_5,W_6, W_7)$ without changing the symmetric capacities. Similar observation applies to all polarization levels $l$, for $1 \le l \le n$. Thus, the number of permutation classes is upper bounded by 
$\frac{N !}{ 2^{1+2+\ldots+\frac{N}{2}}} 
=\frac{N!}{ 2^{1+2+\ldots+2^{n-1}}}
= \frac{N!}{2^{2^{n}-1}}= \frac{N!}{2^{N-1}}.$
\end{IEEEproof}
\begin{remark}
Equivalent classes of permutations can be equally defined in terms of the Battacharaya parameters instead of symmetric capacities, and the statement of Lemma~\ref{symmetry} holds.
\end{remark}
\begin{example}
\label{illustrative_example}
Fig.~\ref{example_4} illustrates the polarization process for $N=4$ channels consisting of two different binary erasures channels (BECs). Based on Lemma~\ref{symmetry}, it can be checked that in this example, the number of permutation classes is 2, as shown in Fig.~\ref{example_4}. For a target rate of $\frac{1}{2}$, the rightmost permutation in Fig.~\ref{example_4} is advantageous as it corresponds to the highest sum-rate of the two best synthetic channels.
\begin{figure} [b]
\centering
\includegraphics[width=1\linewidth]{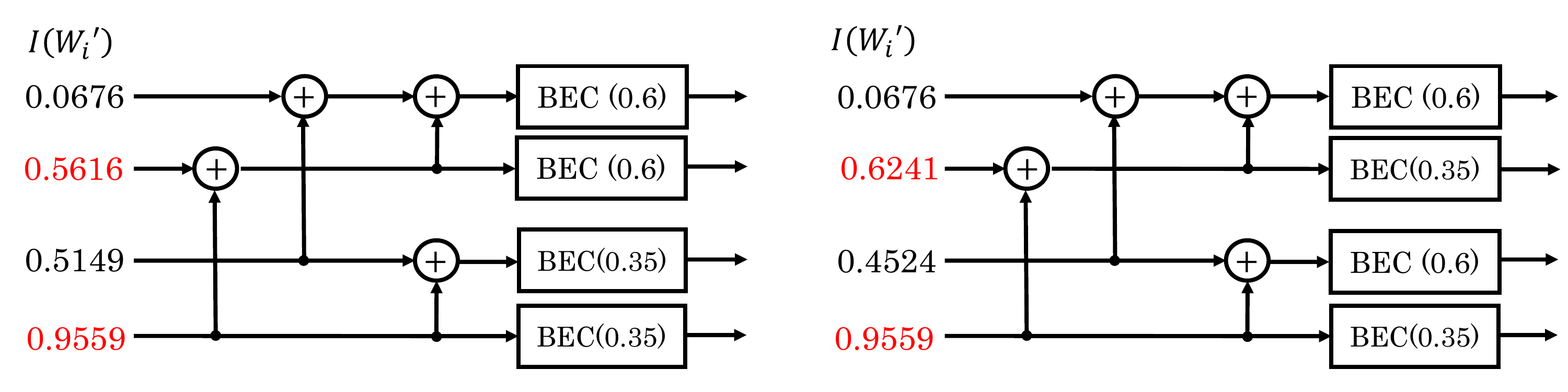}
\caption{Two different ordering of the four BECs, with their synthetic channel rates. The permutation on the left is $\pi=[0,1,2,3]$ and the permutation on the right is $\pi= [0,2,1,3]$.}
\label{example_4}
\end{figure}
\end{example}
Example~\ref{illustrative_example} shows that in the finite blocklength regime, the ordering of the channels matters, depending on the code rate. For $N=4$ BECs with different erasure probabilities, Proposition~\ref{toy_example} determines the best ordering depending on the target rate.
\begin{proposition}
\label{toy_example}
Consider $N=4$ BECs with parameters $\epsilon_i, 1~\le i \le 4$, such that $\epsilon_1 \ge \epsilon_2\ge \epsilon_3\ge \epsilon_4$. Then, the permutation $\pi = [0,3,1,2 ]$ satisfies \eref{permutation_choice}. 
\end{proposition}
Extending the analysis in Proposition~\ref{toy_example} is not tractable. In the following, we consider the bit-reversal permutation defined below. We note that such permutation was used in \cite{niu2013beyond} in the context of rate-compatible punctured polar codes and was shown to yield attractive performance compared to other existing or random puncturing patterns. We explore the performance under bit-reversal permutation through various numerical examples. 
\begin{definition}
We define $\psi$ to be the bit-reversal permutation. For each integer $i \in \{0,\ldots, 2^n-1 \}$, $\psi(i)$ is the integer obtained by reversing the binary representation of $i$. That is, let $i = \sum_{j=1}^n b_j 2^{j-1}$, then $\psi(i) =  \sum_{j=1}^n b_j 2^{n-j }$. As an example, when $N=8$, $\psi= [0 , 4, 2, 6, 1, 5, 3, 7 ]$, as illustrated in Fig.~\ref{PCdiagram}.
\end{definition}
From the proof of Proposition~\ref{toy_example} and Lemma~\ref{symmetry}, when $\epsilon_1~=~ \epsilon_2$ or $\epsilon_3 = \epsilon_4$, then the bit-reversal permutation $\psi$ satisfies \eref{permutation_choice} and is optimal, which coincides with Example~\ref{illustrative_example}. 
\begin{example}
\label{synthetic_example}
We consider $N$  binary symmetric channels (BSCs) with varying cross-over probabilities, linearly spaced and centered at value $p \in \{ 0.05,    0.065,    0.08,    0.095,    0.11\}$, with maximum deviation of $0.045$. We consider a rate $\frac{1}{2}$ and a block size $N=2^{10}$. We evaluate the performance of a regular polar code designed for the \textit{average} BSC channel, with $p_{avg}$ satisfying $N (1-h (p_{avg})) = \sum_{i=0}^{N-1}   (1 - h(p_i)),$ where $h(\cdot)$ is the binary entropy function and $p_i$'s are the cross-over probabilities of the channels. The default ordering of the channels is such that the channels' Bhattacharya parameters are ordered in a descending order (i.e., decreasing order of $p_i$'s). We evaluate the performance with the default ordering (no permutation) and the performance with the bit-reversal permutation. We also evaluate the performance obtained by averaging $200$ random permutations.  For each scenario, we evaluate the frame error rate (FER) and the bit error rate (BER) with non-systematic encoding and systematic encoding~\cite{arikan2011systematic}, for $10^4$ runs. Fig.~\ref{performance_polar} illustrates the results. Similar to~\cite{arikan2011systematic}, we observe that systematic encoding of non-stationary polar codes enhances the BER performance compared to non-systematic encoding, while keeping the FER unchanged. The regular polar code exhibits the worst BER, while the non-stationary polar code under systematic encoding with $\pi = \psi$ performs the best. In particular, the latter scenario outperforms all 200 random permutations for all values of $p$.
\begin{figure} 
\includegraphics[width=1\linewidth]{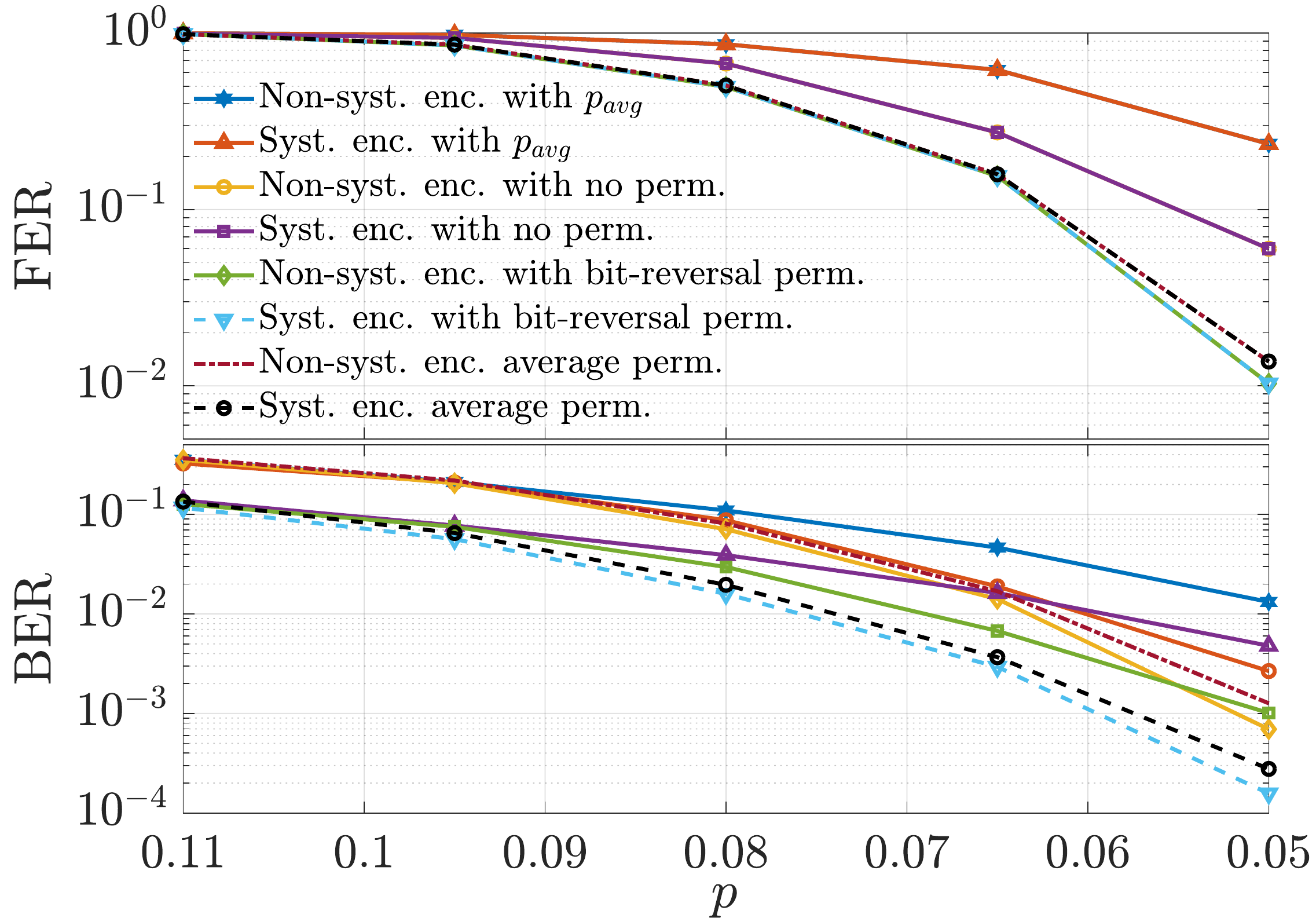}
\caption{Performance evaluation for BSCs with linearly spaced cross-over probabilities. $N= 1024, k = 512$.}
\label{performance_polar}
\vspace{-0.15in}
\end{figure}  
\end{example}
Based on the observation that bit-reversal permutation achieves competitive BER numerically, we choose to apply $\pi =\psi$ for our non-stationary polar codes, so as to mitigate the sneak path problem in crossbar arrays. 
\section{Application to Resistive Crossbar arrays}
\subsection{General Framework}
\label{General_framework}
\begin{figure} 
\centering
\includegraphics[width=0.8\linewidth]{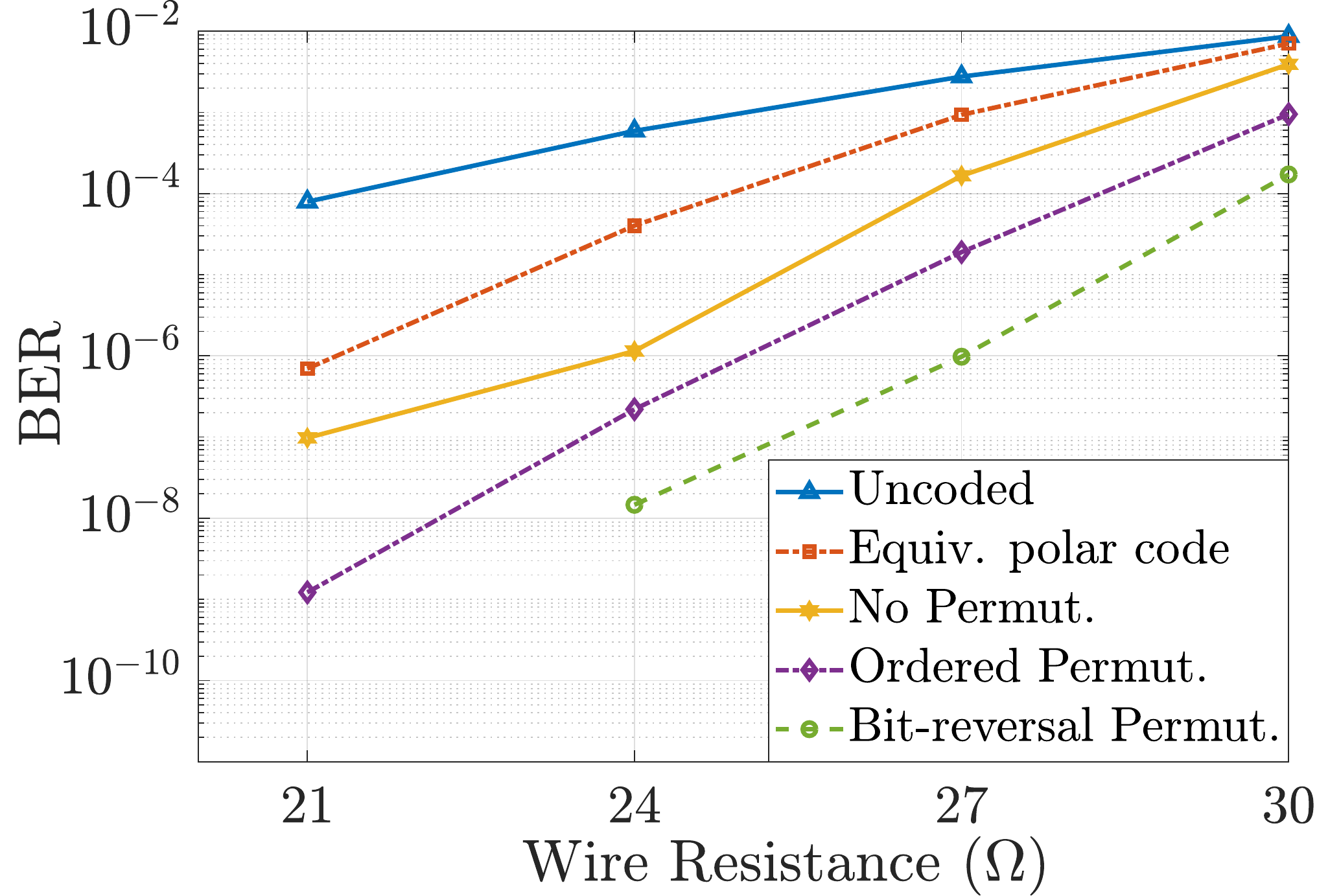}
\caption{Performance evaluation for a $(32 \times 32)$ crossbar array, code rate $k/n=0.8$.}
\label{different_permutations}
\vspace{-0.1in}
\end{figure}

As outlined in the introduction, the crossbar array cells exhibit different reliability levels. For this reason, we propose the application of non-stationary polar codes to address the problem. We apply a two-step approach:
\begin{enumerate}
\item	We first estimate a single detection threshold for each wordline (row) to minimize the overall uncoded BER per word. This transforms the read channels into BSCs. The threshold for each wordline is estimated by generating large training data and then applying a good binary classifier. For instance, we observe that a logistic regression-based classifier gives superior performance in terms of accuracy and speed. 
\item Based on the estimated thresholds in step 1), we estimate the cross-over probabilities of each cell in the array. We can then apply non-stationary polar codes using the cell characterizations.  
\end{enumerate}
In a preliminary version of this work~\cite{zorguipolar}, we proposed to encode each row separately. In this paper, we focus on encoding the entire array. This can be suitable for applications such as archival data and image storage.
Assuming the crossbar array size is $(N_1 \times N_2)$, then, the blocklength is $N=N_1 N_2$. The encoded output symbol $z_{i n + j}, 0 \le i < m, 0 \le j < n $, is stored at the $(i,j)$-th entry in the crossbar array (i.e., we vectorize the array row by row).
Instead of using high-level models for the sneak path problem, such as in~\cite{chen2018coding,ben2019detection}, we use a SPICE-like simulator that is built based on accurate modeling of the resistive crossbar array \cite{fouda2018modeling}. This numerical simulator offers a fast alternative to SPICE simulators while maintaining the same simulation accuracy. In our simulations, the high resistance state (HRS), representing 1, and the low resistance state (LRS), representing 0, are set to $1 M\Omega$ and $1 k \Omega$, respectively.
\begin{example}
In Fig.~\ref{different_permutations}, we simulate the  BER performance of systematic polar codes for four cases: 
(i) equivalent regular polar codes correspond to BSCs with parameter $p_{avg}$, similar to Example~\ref{synthetic_example},
(ii) permutation $\pi = [0, \ldots, N-1 ]$, (iii) permutation $\pi_{\text{ord}}$, and (iv) permutation $\pi= \pi_{\text{ord}}\circ \psi$, where $\pi_{\text{ord}}$ is as defined in Remark \ref{general_order}.  Clearly, the BER permutation under $\pi= \pi_{\text{ord}}\circ \psi$ outperforms the other permutations.
\end{example}
\vspace{-.6cm}
\subsection{Binary Asymmetric Channel Modeling}
In subsection~\ref{General_framework}, the array cells are modeled as BSCs. Analyzing further the uncoded error distribution, as one may infer from Figures \ref{Fig3b}, \ref{Fig3c}, and \ref{Fig3d}, we find that the conditional error distributions under $0$'s and $1$'s are different, i.e., $P(\text{error}|0) \neq P(\text{error}|1)$. Taking this observation into consideration, we model the crossbar array cells as binary asymmetric channels (BACs) and apply the non-stationary polar codes developed in Section~\ref{polar_code_framework}.
\begin{example}
In Fig.~\ref{BAC_vs_BSC}, we compare the systematic BER performance under both BSC and BAC modeling. As expected, the BER performance under the more accurate BAC model is higher, and the gain increases as the wire resistance decreases. 
\begin{figure} 
\centering
\includegraphics[width=0.8\linewidth ]{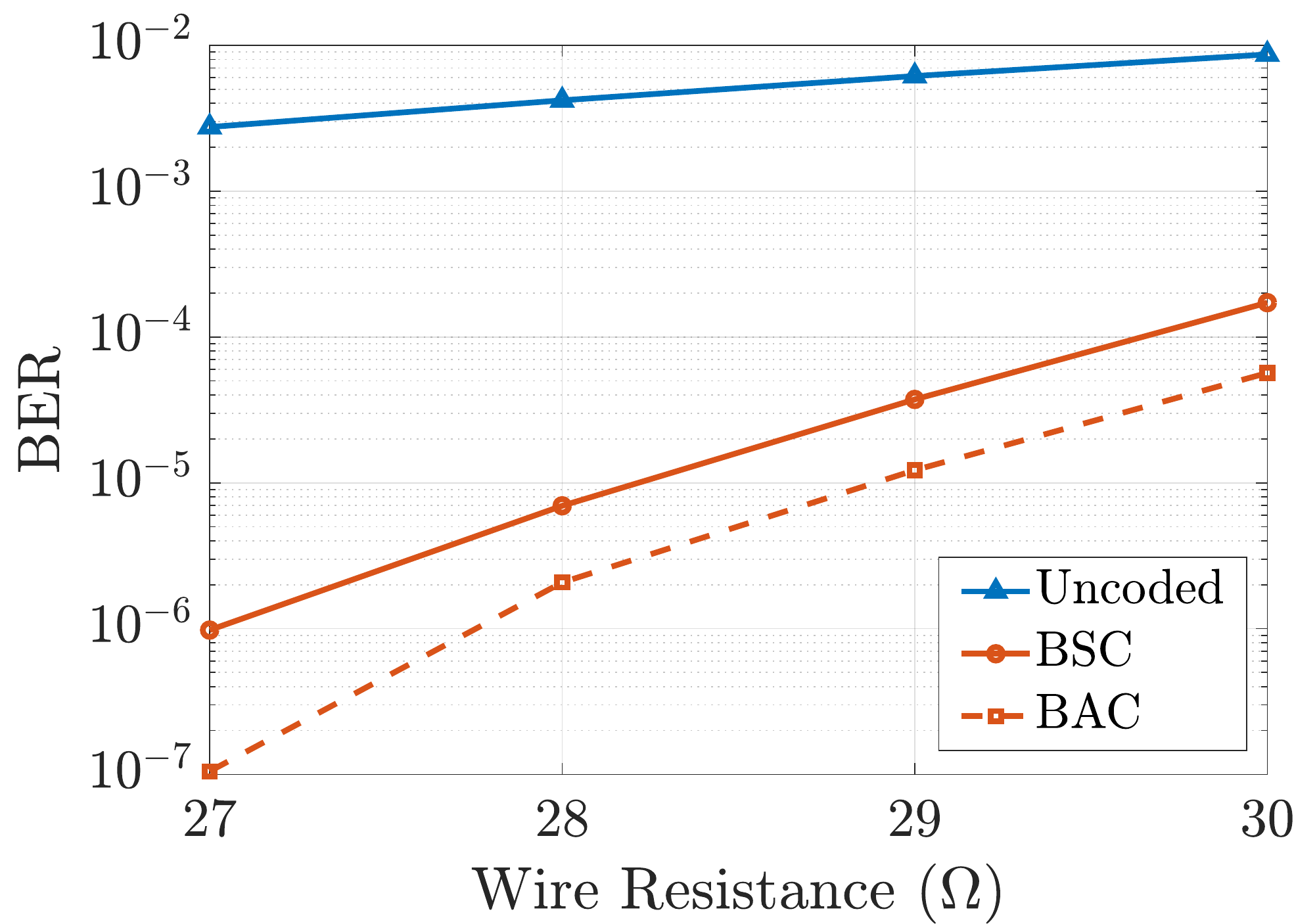}
\caption{BER performance under BSC and BAC modeling for a $(32 \times 32)$ crossbar, code rate $k/n=0.8$.}
\label{BAC_vs_BSC}
\end{figure}
\end{example}
\subsection{Punctured Polar Codes}
In this subsection, we propose a technique that can enhance the BER performance in some scenarios. The sneak paths exist through the cells having low resistances,
causing inter-cell interference \cite{chen2018coding}. Intuitively, having more high resistances in the array helps mitigate the sneak path problem. To leverage this intuition, we investigate the use of a (punctured) polar code of a shorter length, say $N-N_p$, while storing high resistances in the corresponding punctured $N_p$ cells in the array. Clearly, there is a trade-off between two opposite factors: puncturing reduces the number of redundant codeword symbols, hence degrading the performance of polar codes, while high resistances decrease the sneak path effect resulting in fewer read errors. In the following, we investigate the application of the above approach to the crossbar array. 
%

A \textit{punctured} polar code is obtained from the $N$-length parent polar code using a puncturing vector $ \mathbf{w}=[w_1,\ldots, w_N]$, with $w_i \in \{ 0,1\}, i=1,\ldots,N$, where the $0$s imply the punctured positions. We note that the information set $\mathcal{I}$ should be recomputed as in Remark~\ref{computing_information_set}, when we consider puncturing. Punctured polar codes have been investigated by many works, and several efficient puncturing patterns have been proposed in the literature \cite{niu2013beyond}.

In \cite{niu2013beyond}, the authors proposed an empirically good puncturing algorithm, termed quasi-uniform puncturing (QUP). QUP-polar codes were shown through simulations to outperform the performance of
turbo codes in WCDMA (Wideband Code Division Multiple Access) or LTE (Long Term Evolution) wireless communication systems in the large range of code lengths. We adopt QUP as our puncturing pattern and we highlight its advantages below.
\begin{definition}[\cite{niu2013beyond}]
The QUP is described as follows:
\begin{enumerate}
\item	Initialize the vector $\mathbf{w}$ as all ones, and then set the first $N_p$ bits of $\mathbf{w}$ to zeros;
\item	Perform bit-reversal permutation on the vector $\mathbf{w}$ and obtain the puncturing pattern.
\end{enumerate}
\end{definition}
\begin{example}
Let $N=8, N_p=3$. The initial puncturing vector is $\mathbf{w}= [0,0,0,1,1,1,1,1]$. After bit-reversal permutation, the puncturing vector becomes $\mathbf{w}=[0,1,0,1,0,1,1,1]$.
\end{example}
By employing QUP in conjunction with the permutation $\pi_{\text{ord}} \circ \psi$, puncturing $N_p$ positions corresponds to not using the $N_p$ cells with the worst reliability levels for data storage.
By placing high-resistance values (i.e., 1's) in the punctured locations, we increase the frequency of $1$'s in the codeword by $\frac{N_p}{2N}$.
\begin{lemma}
For a punctured polar code length of $N-N_p$, the frequency of $1$'s in the array is given by $\frac{1}{2}+\frac{N_p}{2N}$.
\end{lemma}
\begin{IEEEproof}
Let $i$ be chosen uniformly at random in $\{ 0, \ldots, N-1\}$, and let $z_i$ be the corresponding codeword symbol. Let $\mathcal{B}_p$ be the set of punctured locations. The size of $\mathcal{B}_p$ is $N-N_p$. Then, we write
\begin{align*}
P(z_i = 1) &= P(i \in \mathcal{B}_p) P (z_i = 1 |i \in \mathcal{B}_p )  \nonumber\\
&+ P(i \notin \mathcal{B}_p) P (z_i = 1 |i \notin \mathcal{B}_p ) \\
&= \frac{N_p}{N} + \frac{1}{2} \frac{N-N_p}{N}= \frac{1}{2}+ \frac{ N_p}{2N}.
\end{align*} 
\end{IEEEproof}
\begin{figure} 
\centering
\includegraphics[width=.8\linewidth]{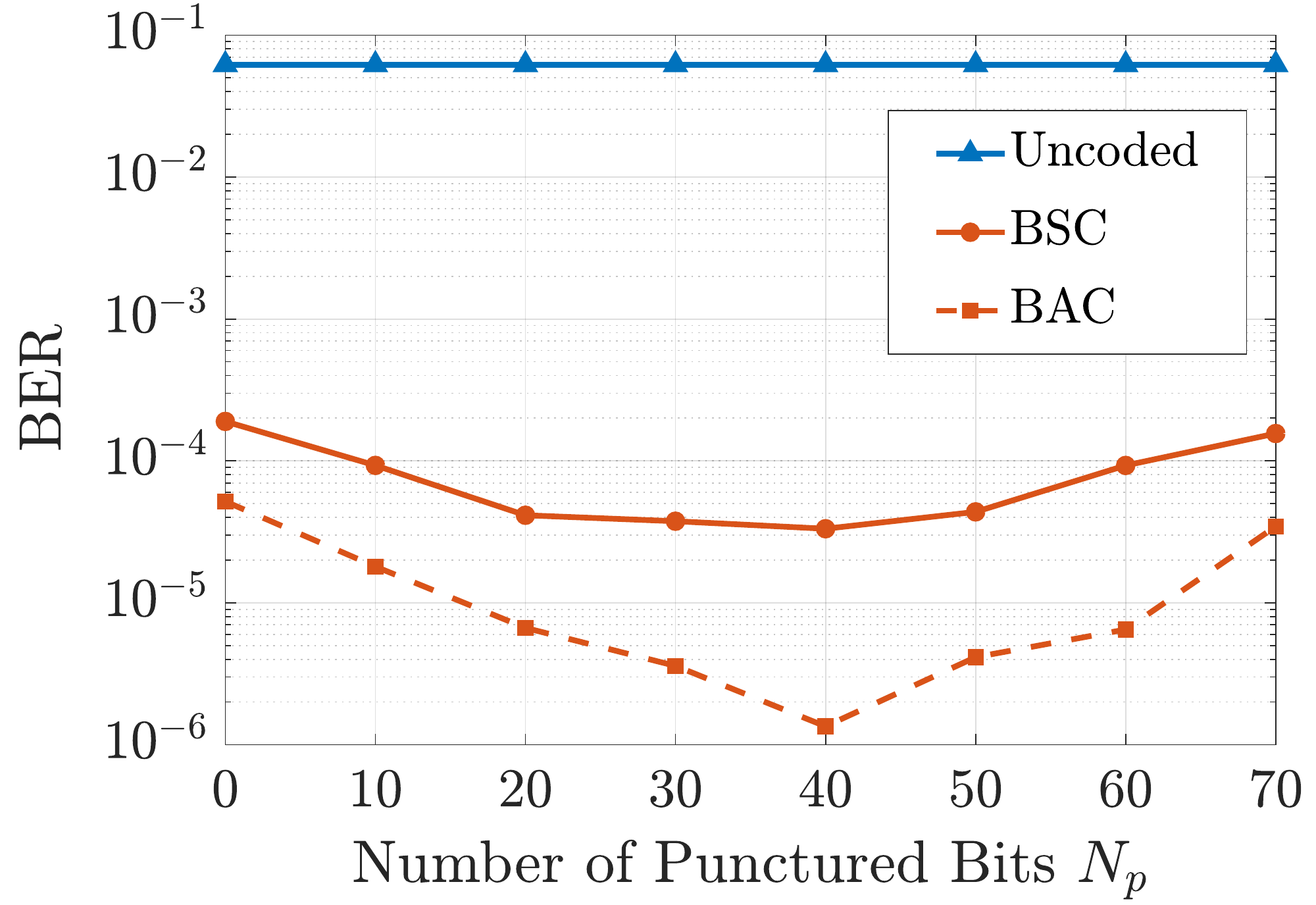}
\caption{Punctured polar encoding over the $(32 \times 32)$ crossbar array for code rate $k/n=0.8$.}
\label{puntured_effect}
\end{figure}
\begin{example}
Fig.~\ref{puntured_effect} illustrates the systematic BER performance of QUP for a $(32 \times 32)$ array with $R_w =35 \Omega, \pi =  \pi_{\text{ord}}\circ \psi$. We observe that the BER decreases as more bits ($N_p$) are punctured and as the frequency of $1$'s increases. This gain is reversed for $N_p > 40$ and the BER increases as $N_p$ and the codeword redundancy decrease. The BER is improved by a factor of $5.7$ for a symmetric channel model and by a factor of $38.5$ for an asymmetric channel model.
\end{example}
\section{Concluding remarks}
\label{concluding_remarks}
In this paper, motivated by the sneak path problem in resistive memories, we studied polar coding over channels with different reliability levels. In particular, we argued that the channels' ordering is important and proposed a channel ordering whose attractive performance was shown numerically.  
We then applied our framework to the sneak path problem in resistive memories. Simulation results on SPICE-like resistive crossbar showed significant bit-error rate performance improvement, especially for low uncoded BER. Additionally, we proposed two approaches to further lower the BER. The first approach relies on a more accurate channel modeling. The second approach consists in biasing the frequency of high-resistance values in the array so as to mitigate the sneak path occurrences. We note that while in this work, we modeled each cell individually as a BSC (or a BAC), the cost of such modeling is amortized by using the same characterization over several crossbar arrays, which makes the model parameters costs justifiable from a practical perspective. Moreover, it is possible to \textit{cluster} multiple cells together in a way to reduce the number of overall crossbar model parameters.

The present work represents another step toward coding for the resistive crossbar arrays. The performance of polar codes can be improved by using more enhanced decoding algorithms (e.g. list-decoding~\cite{tal2015list}), at the expense of higher complexity. Moreover, by the sneak path nature, errors in neighboring cells are not independent. A future direction for research is to investigate enhanced polar decoding algorithms, taking into consideration such correlations. Another avenue for research is to investigate other techniques for biasing the distribution of high-resistance values, along with techniques for coding for asymmetric channels as in~\cite{coding_for_BAC}. 
\bibliographystyle{IEEEtran}
\bibliography{ref22}
\subsection{Proof of Proposition~\ref{toy_example}}
\label{proof_toy_example}
\begin{IEEEproof}
Assume that $\epsilon_1\geq \epsilon_2 \geq \epsilon_3 \geq \epsilon_4$. As for a BEC W, $Z(W)= 1- I(W)$, we can equivalently study the Bhattacharya parameters. By Lemma~\ref{symmetry}, it can be checked that we need only consider 3 different permutations. 
 
\noindent $\bullet$ {Case 1: $ \pi_1= [0, 1, 2, 3 ]$:} using Lemma \ref{single_step_results}, we obtain
\begin{align*}
Z_{\pi_1}^{(0)} &= \epsilon_1 + \epsilon_2 + \epsilon_3 + \epsilon_4 - \epsilon_1 \epsilon_2 - \epsilon_3 \epsilon_4  \\
& 	- (\epsilon_1 + \epsilon_2 - \epsilon_1 \epsilon_2) (\epsilon_3 + \epsilon_4 - \epsilon_3 \epsilon_4), \\
Z_{\pi_1}^{(1)} &= \epsilon_1 \epsilon_2 + \epsilon_3 \epsilon_4 - \epsilon_1 \epsilon_2 \epsilon_3 \epsilon_4, \\
Z_{\pi_1}^{(2)}&=  (\epsilon_1 + \epsilon_2 - \epsilon_1 \epsilon_2) (\epsilon_3 + \epsilon_4 - \epsilon_3 \epsilon_4),\\
Z_{\pi_1}^{(3)} &= \epsilon_1 \epsilon_2 \epsilon_3 \epsilon_4.
\end{align*}


\noindent $\bullet$ {Case 2: $ \pi_2= [0, 2, 1, 3 ]$:} exchanging the roles of $\epsilon_1$ and $\epsilon_2$ in the previous case, we obtain 
\begin{align*}
\nonumber
\begin{bmatrix}
Z_{\pi_2}^{(0)} \\
Z_{\pi_2}^{(1)}\\
Z_{\pi_2}^{(2)} \\
Z_{\pi_2}^{(3)}
\end{bmatrix}=  
\begin{bmatrix}  
Z_{\pi_1}^{(0)}\\
 \epsilon_1 \epsilon_3 + \epsilon_2 \epsilon_4 - \epsilon_1 \epsilon_2 \epsilon_3 \epsilon_4 \\
(\epsilon_1 + \epsilon_3 - \epsilon_1 \epsilon_3) (\epsilon_2 + \epsilon_4 - \epsilon_2 \epsilon_4)\\
Z_{\pi_1}^{(3)}
\end{bmatrix}.
\end{align*}
\noindent $\bullet$ {Case 3: $ \pi_3= [0, 3, 1, 2 ]$:} we obtain
\begin{align*}
\begin{bmatrix}
Z_{\pi_3}^{(0)} \\
Z_{\pi_3}^{(1)}\\
Z_{\pi_3}^{(2)} \\
Z_{\pi_3}^{(3)}
\end{bmatrix}=  
\begin{bmatrix}  
Z_{\pi_1}^{(0)}\\
\epsilon_1 \epsilon_4 + \epsilon_2 \epsilon_3 - \epsilon_1 \epsilon_2 \epsilon_3 \epsilon_4 \\
 (\epsilon_1 + \epsilon_4 - \epsilon_1 \epsilon_4)  (\epsilon_2 + \epsilon_3 - \epsilon_2 \epsilon_3) \\
Z_{\pi_1}^{(3)}
\end{bmatrix}.
\end{align*}
\noindent\textbf{Analysis}: 
Note that $ \sum_{i=0}^{3} Z_{\pi_1}^{(i)} = 
\sum_{i=0}^{3} Z_{\pi_2}^{(i)}
= \sum_{i=0}^{3} Z_{\pi_3}^{(i)}$. Thus, 
$
 Z_{\pi_1}^{(1)} +  Z_{\pi_1}^{(2)} = 
  Z_{\pi_2}^{(1)} +  Z_{\pi_2}^{(2)} =
  Z_{\pi_3}^{(1)} +  Z_{\pi_3}^{(2)}.
$         
Under~$\pi_2$, it follows by Lemma \ref{single_step_results} that $Z_{\pi_2}^{(0)}> Z_{\pi_2}^{(2)}$ and $Z_{\pi_2}^{(1)}>Z_{\pi_2}^{(3)}$. Moreover, we have 

\vspace{-0.3cm}
\begin{small}
\begin{align*}
Z_{\pi_2}^{(0)}-Z_{\pi_2}^{(1)}& = 
(1-\epsilon_1) (1-\epsilon_2) (\epsilon_3+\epsilon_4) +  (\epsilon_1-\epsilon_3) (\epsilon_4-\epsilon_2) \\
&  +(1-\epsilon_3) (1-\epsilon_4) (\epsilon_1 +\epsilon_2) \geq 0, \\
Z_{\pi_2}^{(2)}-Z_{\pi_2}^{(3)}&=  \epsilon_1 \epsilon_3 + \epsilon_2 \epsilon_4 - 2 \epsilon_1 \epsilon_2 \epsilon_3 \epsilon_4  \geq 0.
\end{align*}
\end{small}

\vspace{-0.3cm}
\noindent	It follows that 
$Z_{\pi_2}^{(3)} \le \min( Z_{\pi_2}^{(2)},
Z_{\pi_2}^{(1)} ) \le Z_{\pi_2}^{(0)}$. 
Similarly, one obtains
\begin{align*}
Z_{\pi_1}^{(3)} &\le \min( Z_{\pi_1}^{(2)},
Z_{\pi_1}^{(1)} ) \le Z_{\pi_1}^{(0)}, \\
Z_{\pi_3}^{(3)} &\le \min( Z_{\pi_3}^{(2)},
Z_{\pi_3}^{(1)} ) \le Z_{\pi_3}^{(0)}.
\end{align*}
Given the exact expressions of the Battacharaya parameters, we can determine the permutation satisfying~\eref{permutation_choice}, depending on the size of $\mathcal{I}$, i.e., depending on the code rate. Clearly, except for the rate $\frac{1}{2}$, all permutations satisfy~\eref{permutation_choice}. For example, if $R=1/4$, we choose the same best channel corresponding to $Z_{\pi_1}^{(3)}$ for all permutations. Thus, we only need to analyze the rate $\frac{1}{2}, i.e., |\mathcal{A}| = 2$.
We first compare permutations $\pi_2$ and $\pi_3$. 
\begin{align*}
Z_{\pi_2}^{(1)} - Z_{\pi_3}^{(1)} &=
Z_{\pi_3}^{(2)} - Z_{\pi_2}^{(2)} =
 (\epsilon_1 - \epsilon_2) (\epsilon_3 - \epsilon_4) \geq 0, \\
 Z_{\pi_3}^{(1)} - Z_{\pi_2}^{(2)} &= Z_{\pi_2}^{(1)} - Z_{\pi_3}^{(2)} \\
&= \epsilon_1 \epsilon_2 \epsilon_3 - \epsilon_3 \epsilon_4 - \epsilon_1 \epsilon_2   + \epsilon_1 \epsilon_2 \epsilon_4 + \epsilon_1  \epsilon_3 \epsilon_4
\\ &   + \epsilon_2 \epsilon_3 \epsilon_4 - 2 \epsilon_1 \epsilon_2 \epsilon_3 \epsilon_4 \\
&=- \epsilon_3 \epsilon_4 ( 1-\epsilon_1 ) ( 1-\epsilon_2 )- \epsilon_1 \epsilon_2 ( 1-\epsilon_3 ) ( 1-\epsilon_4 ) \\
&\le 0.
\end{align*}
It follows that  $Z_{\pi_3}^{(1)}  \le \min ( Z_{\pi_2}^{(1)},  Z_{\pi_2}^{(2)} ) \le Z_{\pi_3}^{(2)}$. Similarly, one obtains $Z_{\pi_3}^{(1)} \le \min ( Z_{\pi_1}^{(1)},  Z_{\pi_1}^{(2)} )
\le Z_{\pi_3}^{(2)}$.
Therefore, $Z_{\pi_3}^{(1)}$ is the second best synthetic channel across all 3 permutations, which implies that $\pi_3$ satisfies \eref{permutation_choice}. 
\end{IEEEproof}
\end{document}